\documentclass[12pt,preprint]{aastex}

\newcommand{\dt}{$\Delta\theta$\ }
\newcommand{\ddt}{$\Delta\theta$}
\shorttitle{The alignment of the polarization of HAe/Be stars with the interstellar magnetic field}
\shortauthors{Rodrigues et al.}

\begin{document}


\title{The alignment of the polarization of Herbig Ae/Be stars with the interstellar magnetic field\footnote{Based on observations made at the Observat\'orio do Pico dos Dias, Brazil, operated by the Laborat\'orio Nacional de Astrof\'\i sica.}}


\author{Cl\'audia V. Rodrigues\altaffilmark{2}, 
Mar\'\i lia J. Sartori\altaffilmark{3}, 
Jane Gregorio-Hetem\altaffilmark{4},
A. M\'ario Magalh\~aes\altaffilmark{4}}


\altaffiltext{2}{Instituto Nacional de Pesquisas Espaciais/MCT --
Av. dos Astronautas, 1758 -- 12227-010 - S\~ao Jos\'e dos
Campos - SP -- Brazil; claudiavr@das.inpe.br}

\altaffiltext{3}{Laborat\'orio Nacional de Astrof\i sica/MCT -- 37504-364 - Itajub\'a - MG - Brazil}

\altaffiltext{4}{Instituto de Astronomia, Geof\'\i sica e Ci\^encias Atmosf\'ericas/Un. S\~ao Paulo -- R. do Mat\~ao, 1225 -- 05508-900 - S\~ao Paulo - SP -- Brazil}


\begin{abstract}

We present a study of the correlation between the direction of the symmetry axis of the circumstellar material around intermediate mass young stellar objects and that of the interstellar magnetic field. We use CCD polarimetric data on 100 Herbig Ae/Be stars. A large number of them shows intrinsic polarization, which indicates that their circumstellar envelopes are not spherical. The interstellar magnetic field direction is estimated from the polarization of field stars. There is an alignment between the position angle of the Herbig Ae/Be star polarization and that of the field stars for the most polarized objects. This may be an evidence that the ambient interstellar magnetic field plays a role in shaping the circumstellar material around young stars of intermediate mass and/or in defining their angular momentum axis.




\end{abstract}





\section{Introduction}
The role of the magnetic field in the star formation processes is a longstanding problem in astrophysics. The magnetic field is proposed to act in the support of molecular clouds against the gravitational force by some authors \citep[e.g.,][]{mou99}. A different view is that these clouds are not stable and exist as ephemeral structures. In this case, the turbulence drives the interstellar medium large scale structure \citep[e.g.,][]{mo07}. It is possible to address this issue by the search of evidences about the importance of interstellar magnetic field in the star forming process. From an observational point of view, many works have searched for a correlation between the direction of the interstellar magnetic field and the geometry of young stellar objects (YSOs) traced by disk axis, outflow direction, or observed polarization. We cite two of them. \citet{tam89} have found a correlation between the interstellar magnetic field direction and the infrared polarization angle in a sample of 47 YSOs in Taurus-Auriga molecular cloud. This sample is dominated by T Tauri stars. Recently, \citet{men04} have studied  37 T Tauri stars in the same region and found no correlation between the local magnetic field and the geometry of the YSOs.

Herbig Ae/Be objects (HAeBe) are pre-main sequence stars, analogue to T Tauri stars, but of intermediate mass. Lists of HAeBe stars can be found in the compilation of \citet{the94} and in The Pico dos Dias Survey (PDS), a search for T Tauri stars based on IRAS colors \citep{gre92,tor95}. In spite of the focus on low mass YSOs, PDS has also found around a hundred of HAeBe candidates \citep{vie03,tor99}. In this work, we revisited the issue of alignment of the interstellar magnetic field with the YSO geometry using a large sample of HAeBe objects. The study of the polarization in the context of the circumstellar material properties will be done elsewhere (Sartori M. et al., in preparation). In Section \ref{sec_obs}, we describe the acquisition and reduction of the polarimetric data and the technique to calculate the interstellar and intrinsic stellar polarization. The results and discussion are presented in Section \ref{sec-res}. In the last section, we summarize our findings.

\section{Observations}
\label{sec_obs}


We obtained polarimetric data on 102 fields containing probable HAeBe stars selected from \citet{the94}, \citet{tor99}, and \citet{vie03}. The observations were done with the 0.60-m Boller \& Chivens telescope at the Observat\'orio do Pico dos Dias, Brazil, operated by the Laborat\'orio Nacional de Astrof\'\i sica, Brazil, from 1998 to 2002. We used a CCD camera modified by the polarimetric module described in \citet{mag96}. The used detector is a SITe back-illuminated CCD, $1024 \times 1024$ pixels. This combination of telescope and instrumentation results in a field-of-view of 10\farcm5 $\times$ 10\farcm5 (1 pixel = 0\farcs62). The data were taken using a $V$ filter. We have collected eight images of each field. Table \ref{tab_pol} lists the observation date and the integration time (for one image) for each field. The reduction followed the standard steps of differential photometry using the IRAF facility\footnote{IRAF is distributed by National Optical Astronomy Observatories, which is operated by the Association of Universities for Research in Astronomy, Inc., under contract with the National Science Foundation.} and the package {\sc pccdpack} \citep{per00,per02}. Polarized standard stars were observed to convert the instrumental position angle to the equatorial reference frame. Unpolarized standard star measurements were consistent with zero within the errors and hence no corrections for instrumental polarization were applied to the data. Measurements using a Glan filter, which provide the efficiency of the instrument, indicate that no correction is needed considering the instrumental precision. The observed polarization data are presented in Table \ref{tab_pol}. It contains the 102 program stars plus 2 confirmed post-AGB that contaminate the PDS sample and are not included in the analysis.

The observed polarization of a YSO is usually composed by two components: an intrinsic polarization plus an interstellar polarization component. The intrinsic component is produced by the scattering of the central source emission off the circumstellar material. In the path from the object to the observer, the interstellar medium introduces a foreground polarization. These two polarizations are combined vectorially to produce the observed polarization. Hence, to obtain the intrinsic polarization one must estimate the foreground value to be subtracted from the observed polarization. 

The striking majority of stars in the sky is either intrinsically unpolarized or has small, uncorrelated intrinsic polarization. The stellar field is hence dominated by objects presenting only the interstellar component. For our purposes we may then estimate the foreground polarization towards the object of interest by a weighted average of the polarization of the field stars with a good signal-to-noise ratio ($P > 3  \sigma_P $). The average was done from the Stokes parameters Q and U. The number of objects in each of our fields vary from 3 to more than 1000. Such a high number of field objects having angular distances to the HAeBe smaller than 5\arcmin\ gives us confidence that this technique to estimate the interstellar component is reliable. It is also probably statistically better than the estimates in previous works in which the foreground objects are located angularly farther from the YSO and in smaller numbers due to the use of photomultiplier as the detector. 

The foreground and intrinsic polarizations, as well as the number of field objects used in the estimate of the interstellar polarization, are shown in Table \ref{tab_pol}. The polarization of each foreground star for each field will be available as a Vizier catalog as well as the polarization vectors superposed on an optical image. Two objects, HD~23302 and HD~23480, are too bright and there is no other object in the image with sufficient signal-to-noise to enable an estimate to the foreground polarization. These two objects are not included in the following analysis. Our final sample analysis is therefore composed by 100 objects.

\section{Results and discussion}
\label{sec-res}

The interstellar polarization is caused by aligned aspherical grains that produce the dichroism of the interstellar medium. The alignment mechanism is not yet completely understood. The classical mechanism is based on paramagnetic dissipation by rotating grains with superparamagnetic inclusions \citep{dav51,jon67}. However, this mechanism may be not efficient enough. Recently, a promising mechanism based on radiative torques has been proposed \citep{lh07}. In both cases, however, the direction of the magnetic field projected in the plane of the sky can be traced by the position angle of the optical interstellar polarization. A recent review on grain alignment can be found at \citet{laz07}.

The intrinsic polarization position angle is related to the axis of symmetry of the HAeBe envelope. It depends on many factors but we can say in a simple way that in optically thin envelope the polarization is parallel to symmetry axis, and in optically thick case it is perpendicular. A proper understanding of this issue is obtained by the modelling of the scattering of the central source light in the YSO circumstellar material \citep[e.g.,][]{bm77,bm90,wh93}.

Is the geometry of HAeBe stars in our sample related to the direction of the interstellar magnetic field? We try to answer such a question by checking whether the position angles of the intrinsic polarization and the surrounding interstellar polarization are correlated. For this purpose, we define \dt as the difference between the intrinsic and foreground polarization directions. \dt runs from 0 to 90\degr. 

\begin{figure}
\includegraphics[height=5in,angle=270]{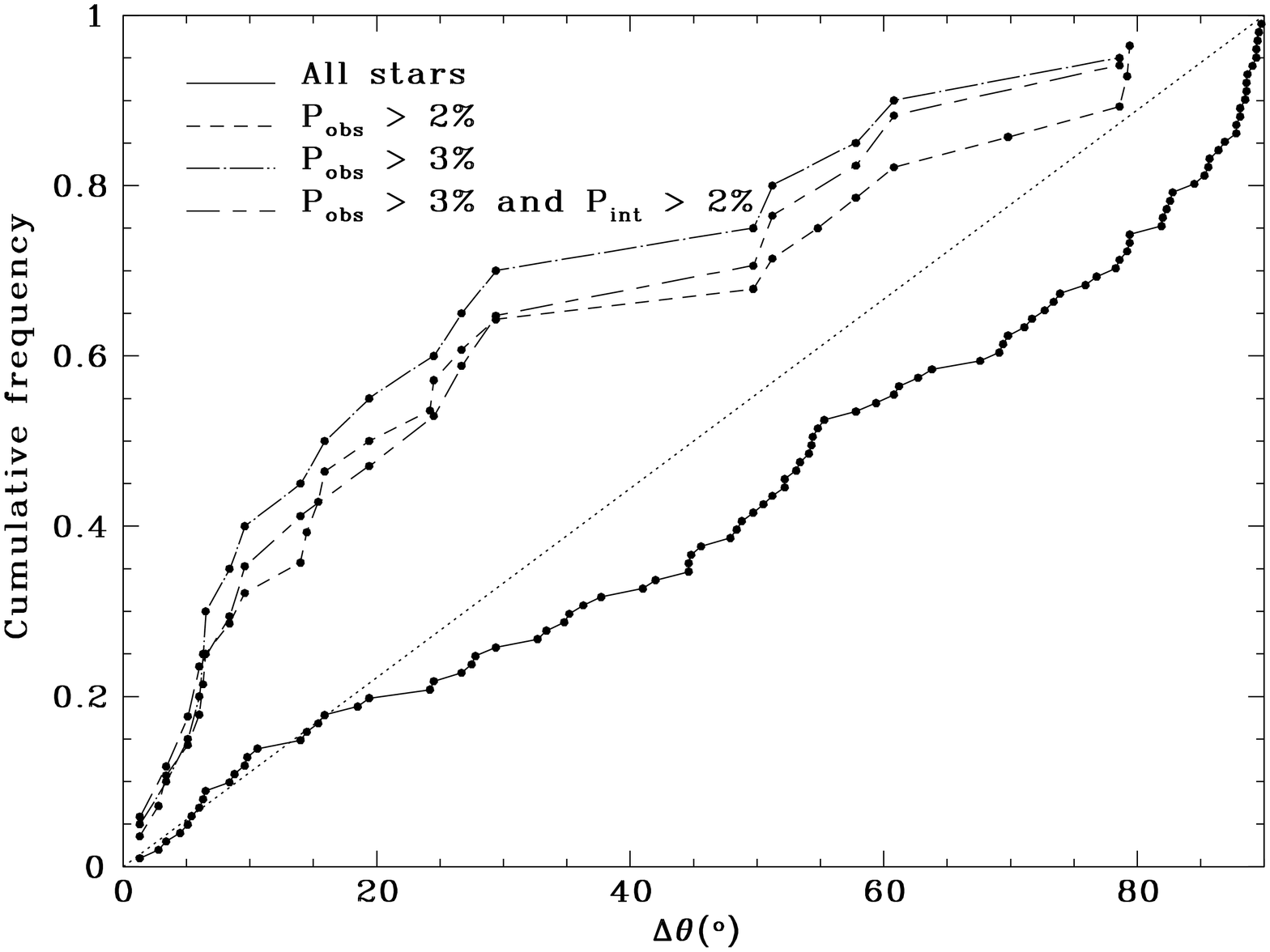}
\caption{Cumulative frequency distributions of the difference between the intrinsic and interstellar polarization angle, \ddt, for our HAeBe sample and sub-samples. See text por details.}
\label{fig_dtheta}
\end{figure}

Figure \ref{fig_dtheta} (solid line) shows the cumulative histogram of \dt for the 100 objects of our sample. The dotted straight line represents the behavior of an uniform distribution. To compare quantitatively both distributions, we used the Kuiper statistic \citep[e.g.,][]{pal04}. It is a modification of the Kolmogorov-Smirnov test and is appropriate for cyclic quantities such as \ddt. Figure \ref{fig_dtheta} shows that the sample as a whole is obviously not uniform. Indeed, the resulting Kuiper statistic has a probability of only 1.3\%, so the hypothesis that our observed distribution is uniform can be discarded. In addition, as the observed curve stays {\it below} that of the uniform distribution, it means that \dt concentrates around 90\degr. 

An inspection of the data shows that this behavior is caused by objects with small values of {\it observed} polarization. In this case, the intrinsic polarization has the same modulus as the foreground polarization but is perpendicular to it; this causes the concentration of \dt near 90\degr. This probably arises as a result of two factors. The HAeBe object may be nearer to us than most of the field stars. In that case, the foreground polarization must be negligible and the observed polarization should have not been corrected by a foreground component. In addition, if the {\it real} intrinsic polarization is undetected given the errors, our estimate of the intrinsic polarization is wrong: it simply reflects the foreground polarization rotated by 90\degr. We may add that a situation in which the interstellar polarization would be perpendicular and have the same modulus of the intrinsic component is less likely. As a result of this discussion, the concentration of \dt around 90\degr\ may carry an observational bias. 

To circumvent such problems, we also built the cumulative histogram considering only objects having the observed polarization larger than 3\% (Figure \ref{fig_dtheta}, dot-dashed line): these are 19 in number. These objects are less affected by the foreground correction, since they must have a larger contribution from the intrinsic polarization to the observed value. In addition, the polarization in all these cases has a high signal-to-noise ratio and hence a well determined interstellar component. In this sample, the Kuiper statistic for \dt has a probability smaller than 2\%. Using (the 27) objects with polarization larger than 2\% (Figure \ref{fig_dtheta}, dashed line), the statistic has this same probability. We also note that, as shown in Figure \ref{fig_dtheta}, these two samples now present \dt clustered around zero.

Star forming regions are dense portions of the ISM. Consequently they usually present high values of interstellar extinction and polarization. Because of that, we have also checked whether the above results might be biased by cases in which the observed polarization is dominated by the interstellar component. Figure \ref{fig_dtheta} (short dash - long dash line) shows the cumulative distribution considering the objects with the observed polarization larger than 3\% and the intrinsic polarization larger than 2\% (16 objects), a sub-sample that excludes cases where the interstellar component is the predominant one. This constraint does not modify significantly the histogram and the resulting Kuiper statistic has a probability smaller than 12\%. 

We concluded that the observed distribution of \dt for the sub-sample of objects presenting high signal-to-noise measurements and reliable values of intrinsic polarization is non uniform with a clear excess of objects with \dt around zero. This result suggests that the polarization of HAeBe stars has a tendency to be aligned with the ambient interstellar magnetic field. It can be interpreted as an indication that the magnetic field of the material that collapsed to form the star can play a role in defining the YSO geometry and/or the symmetry axis of the envelope.

Our findings contrast with the work of \citet{men04} which does not show an alignment between the YSO axis and the interstellar magnetic field using a sample of T Tauri stars. However, previous works, using samples of less evolved objects, indicate an alignment \citep{kob78,dyc79,hec81,hod84,coh84,str87}. A possible solution for this discrepancy is that the alignment may be more easily traced in the less evolved, low mass YSO. This occurs because, during its slow pre-main sequence evolution, an object can move away from its birth place or present a rotation of its axis direction. From a point of view of the YSO mass, HAeBe stars evolve faster than T Tauri stars, which could make the observation of the alignment more probable in the higher mass group. We note that in our sample objects with P greater than 3\% also have larger mid-infrared excess (Sartori et al., in preparation), putting these objects in an early evolutionary stage. 

Some observational arguments have been recently put forward in favor of a fossil origin of the magnetic fields in Ap/Bp stars \citep[e.g.,][]{wad09}. This hypothesis assumes that the interstellar magnetic field present in the cloud that originates the star is amplified along the star evolution and is present in the main sequence stage.  This is another piece of evidence that the formed (proto)stellar object can have a memory of the interstellar magnetic field of the parent cloud and its direction, as indicated by our results.

\section{Conclusions}

We present the results of optical CCD polarimetry of a sample of 102 fields containing HAeBe stars. The direction of the intrinsic polarization of the YSO, and hence their envelope axis, shows a correlation with interstellar magnetic field direction for the sample as a whole. This result may be an observational bias, as discussed in the text. Sub-samples of the more polarized objects present a statistically significant tendency to have the YSO polarization aligned with the interstellar magnetic field. This indicates that the geometry of HAeBe objects retained a memory of the interstellar magnetic field. 

\acknowledgments

We acknowledge the use of the NASA's Astrophysics Data System Service and the SIMBAD database, operated at CDS, Strasbourg, France. This work was partially supported by Fapesp (CVR and AMM: Proc. 2001/12589-1; JGH: Procs. 2001/09018-2 and 2005/00397-1) and CNPq (AMM).

This is an author-created, un-copyedited version of an article accepted for publication in The Astrophysical Journal. IOP Publishing Ltd is not responsible for any errors or omissions in this version of the manuscript or any version derived from it.



{\it Facilities:} \facility{LNA:BC0.6m ()}




\clearpage

\clearpage

\begin{deluxetable}{lrccrrrrrrrrrrc}
\tabletypesize{\scriptsize} 
\rotate
\tablecolumns{15} 
\tablewidth{0pc} 
\tablecaption{Polarimetry of HAeBe objects} 
\tablehead{ 
\colhead{}  & \colhead{} & \colhead{} & \colhead{} & 
\multicolumn{3}{c}{Observed}  & 
\multicolumn{4}{c}{Foreground}   & 
\multicolumn{3}{c}{Intrinsic} \\ 
\cline{5-7} \cline{8-10} \cline{11-14} \\ 
\colhead{Object} & \colhead{V\tablenotemark{1}} & \colhead{Exposure} & \colhead{UT Date}   & \colhead{$P$}   & 
\colhead{$\sigma_P$}    & \colhead{$PA$} & 
\colhead{\# of} &
\colhead{$P$}    & \colhead{$\sigma_P$}   & \colhead{$PA$}    
& \colhead{$P$} & \colhead{$\sigma_P$} & \colhead{$PA$}
& \colhead{Ref.\tablenotemark{2}} \\
\colhead{} &\colhead{$(mag)$} &\colhead{(s)} &\colhead{} &
\colhead{(\%)}   & \colhead{(\%)}    & \colhead{(deg)} & 
\colhead{objects} &
\colhead{(\%)}    & \colhead{(\%)}   & \colhead{(deg)} 
& \colhead{(\%)} & \colhead{(\%)} & \colhead{(deg)}}
\startdata 
PDS 002	& 	10.9 &  100 & 1998 Nov 25 & 0.128	&	0.045	&	135.4	&	12	&	0.190	&	0.022	&	107.0	&	0.161	&	0.050	&	176.1	&	1	\\
PDS 004	&	10.7 &  120 & 1998 Nov 25 & 1.018	&	0.033	&	45.8	&	48	&	0.557	&	0.020	&	49.8	&	0.473	&	0.039	&	41.0	&	1	\\
HD 23302\tablenotemark{3} & 	3.7 &	5 & 1999 Jul 28 & 0.238	&	0.038	&	113.8	&	0	&	-	&	-	&	-	&	-	&	-	&	-	&	3	\\
HD 23480\tablenotemark{3} & 4.2	& 7 & 1999 Jul 28 &	0.507	&	0.033	&	141.2	&	0	&	-	&	-	&	-	&	-	&	-	&	-	&	3	\\
PDS 168	& 17.3 & 300 & 1999 Jul 29 &	8.377	&	0.278	&	62.0	&	33	&	2.544	&	0.015	&	62.9	&	5.835	&	0.278	&	61.6	&	1	\\
PDS 172	& 7.6 &  30 & 1998 Nov 24 &	0.354	&	0.013	&	52.1	&	30	&	0.126	&	0.070	&	127.8	&	0.469	&	0.071	&	48.4	&	 1,2	\\
PDS 176	& 9.8 &  100 & 1998 Nov 24 &	0.141	&	0.021	&	64.4	&	33	&	0.173	&	0.019	&	17.6	&	0.230	&	0.028	&	88.7	&	 1,2	\\
PDS 110	& 10.4 &  200 & 1998 Nov 24 &	0.366	&	0.066	&	27.6	&	21	&	0.367	&	0.024	&	117.9	&	0.733	&	0.070	&	27.7	&	3	\\
PDS 178	& 7.8 &  25 & 1998 Nov 24 &	0.120	&	0.021	&	140.6	&	14	&	2.369	&	0.094	&	83.3	&	2.421	&	0.096	&	172.0	&	1	\\
PDS 179	& 9.7 &  90 & 1998 Nov 25 &	0.336	&	0.027	&	80.0	&	61	&	0.260	&	0.015	&	87.8	&	0.111	&	0.031	&	60.3	&	1	\\
PDS 180	& 10.1 &  120 & 1999 Apr 08 &	0.266	&	0.026	&	58.9	&	25	&	0.188	&	0.011	&	90.6	&	0.248	&	0.028	&	37.5	&	 1,2	\\
PDS 114	& 10.0 &  90 & 1998 Nov 25 &	0.161	&	0.032	&	35.5	&	89	&	0.060	&	0.022	&	10.5	&	0.131	&	0.039	&	45.7	&	1	\\
PDS 184	& 10.1 &  120 & 1999 Apr 08 &	0.421	&	0.017	&	118.5	&	65	&	0.616	&	0.013	&	122.4	&	0.207	&	0.021	&	40.4	&	1	\\
PDS 183	& 8.3 &  30 & 1998 Nov 24 &	0.368	&	0.011	&	41.9	&	21	&	1.849	&	0.084	&	88.1	&	1.900	&	0.085	&	3.6	&	 1,2	\\
PDS 185	& 6.5 & 2 & 2002 Oct 19 &	0.455	&	0.043	&	179.0	&	22	&	0.972	&	0.099	&	168.8	&	0.568	&	0.108	&	70.7	&	1	\\
PDS 190	& 9.3 &  90 & 1998 Nov 25 & 0.390	&	0.149	&	75.5	&	20	&	0.319	&	0.071	&	104.1	&	0.345	&	0.165	&	50.0	&	1	\\
PDS 191	& 8.9 &  40 & 1999 Apr 08 &	0.173	&	0.007	&	75.4	&	5	&	0.153	&	0.011	&	41.4	&	0.183	&	0.013	&	100.8	&	 1,2	\\
PDS 192	& 9.9 &	40 &  2002 Oct 19 & 0.359	&	0.026	&	46.8	&	93	&	0.081	&	0.033	&	18.1	&	0.323	&	0.042	&	52.9	&	1	\\
PDS 193	& 13.9 & 400 & 1998 Nov 25 &	2.183	&	0.114	&	86.3	&	9	&	0.087	&	0.057	&	17.2	&	2.249	&	0.127	&	87.0	&	 1,2	\\
PDS 194	& 9.8 &  100 & 1999 Apr 08 &	0.300	&	0.023	&	31.1	&	6	&	2.035	&	0.304	&	123.9	&	2.334	&	0.305	&	33.5	&	 1,2	\\
PDS 016	& 9.0 &  40 & 1998 Nov 24 &	0.269	&	0.036	&	12.3	&	29	&	0.227	&	0.033	&	176.4	&	0.142	&	0.049	&	41.0	&	1	\\
PDS 201	& 8.9 &  50 & 1998 Nov 24 &	0.573	&	0.037	&	58.4	&	41	&	0.202	&	0.011	&	35.0	&	0.459	&	0.039	&	67.7	&	1	\\
PDS 019	& 13.9 & 200 & 1999 Apr 08 &	0.274	&	0.033	&	129.0	&	37	&	0.259	&	0.013	&	109.9	&	0.175	&	0.035	&	162.1	&	1	\\
PDS 020	& 10.6 &  90 & 1998 Nov 25 &	0.571	&	0.032	&	165.7	&	146	&	1.171	&	0.011	&	161.1	&	0.614	&	0.034	&	66.8	&	1	\\
PDS 021	& 10.4 &  100 & 1998 Nov 25 &	1.598	&	0.022	&	29.7	&	41	&	0.738	&	0.013	&	32.6	&	0.867	&	0.026	&	27.2	&	1	\\
PDS 022	& 10.2 & 100 & 1998 Nov 25 &	0.201	&	0.018	&	102.0	&	82	&	0.083	&	0.010	&	134.6	&	0.182	&	0.021	&	89.8	&	1	\\
PDS 225	& 6.9 & 2 & 1999 Apr 11 &	0.974	&	0.044	&	26.3	&	15	&	0.681	&	0.219	&	172.7	&	0.948	&	0.223	&	47.0	&	1	\\
HD 51585 & 11.1	& 80 & 1999 Apr 11 &	0.522	&	0.045	&	43.3	&	207	&	0.183	&	0.018	&	12.5	&	0.464	&	0.048	&	53.5	&	3	\\
PDS 241	& 12.1 & 120 & 1999 Apr 11 &	3.672	&	0.037	&	104.8	&	315	&	0.324	&	0.009	&	127.4	&	3.451	&	0.038	&	102.9	&	1	\\
PDS 249	& 14.2 & 300 & 1999 Apr 08 &	1.703	&	0.190	&	54.6	&	204	&	0.545	&	0.007	&	147.1	&	2.246	&	0.190	&	55.2	&	1	\\
PDS 272	& 9.8 & 40 & 1999 Jan 22 & 0.130	&	0.026	&	109.6	&	129	&	0.183	&	0.019	&	83.1	&	0.147	&	0.032	&	150.7	&	1	\\
PDS 277	& 10.0 & 45 & 1999 Jan 22 &	0.040	&	0.049	&	97.0	&	178	&	0.537	&	0.014	&	51.2	&	0.540	&	0.051	&	139.0	&	1	\\
PDS 031	& 8.5 & 8 & 1999 Apr 11 &	0.138	&	0.043	&	82.1	&	77	&	0.687	&	0.035	&	87.8	&	0.552	&	0.055	&	179.2	&	1	\\
HBC 563	& 14.2 & 200 & 1999 Jan 22 &	3.307	&	0.241	&	177.7	&	54	&	0.067	&	0.021	&	26.6	&	3.272	&	0.242	&	177.2	&	2	\\
PDS 033	& 12.3 & 80 & 1999 Jan 22 & 0.862	&	0.071	&	76.7	&	82	&	0.034	&	0.025	&	121.1	&	0.862	&	0.075	&	75.5	&	1	\\
PDS 034	& 14.0 & 400 & 1999 Jan 23 &	2.799	&	0.418	&	117.6	&	229	&	0.515	&	0.006	&	104.9	&	2.344	&	0.418	&	120.3	&	1	\\
HD 76534 & 8.0	& 10 & 1999 Apr 11 &	0.466	&	0.017	&	127.4	&	29	&	2.813	&	0.041	&	76.3	&	2.947	&	0.044	&	161.9	&	2	\\
PDS 281	& 8.9 & 24 & 1999 Jan 23 & 1.375	&	0.039	&	160.2	&	28	&	0.067	&	0.032	&	106.2	&	1.397	&	0.050	&	161.5	&	1	\\
PDS 286	& 12.2 & 250 & 1999 Jan 18 &	8.217	&	0.060	&	171.8	&	58	&	1.623	&	0.019	&	178.5	&	6.649	&	0.063	&	170.1	&	1	\\
PDS 290	& 14.5 & 300 & 1999 Apr 11 &	2.408	&	0.227	&	149.8	&	234	&	0.669	&	0.012	&	147.8	&	1.741	&	0.227	&	150.6	&	1	\\
GSC 8593-2802 &	12.0 & 300 & 1999 Jan 22 &	1.946	&	0.058	&	115.8	&	448	&	1.146	&	0.007	&	132.8	&	1.184	&	0.058	&	99.4	&	1	\\
HD 85567 & 8.6 & 14 & 1999 Jan 21 &	0.478	&	0.035	&	105.7	&	174	&	0.715	&	0.035	&	116.1	&	0.317	&	0.049	&	42.2	&	2	\\
PDS 303	& 9.3 & 25 & 1999 Jan 21 & 0.579	&	0.044	&	125.3	&	273	&	0.587	&	0.018	&	116.8	&	0.173	&	0.048	&	167.3	&	1	\\
PDS 037	& 13.5 & 300 & 1999 Jan 18 &	3.253	&	0.104	&	120.1	&	217	&	0.530	&	0.010	&	131.9	&	2.775	&	0.104	&	117.9	&	1	\\
PDS 315	& 10.9 & 80 & 1999 Jan 21 &	2.141	&	0.034	&	158.6	&	681	&	0.931	&	0.007	&	173.3	&	1.406	&	0.035	&	149.1	&	1	\\
GSC 8618-2363 & 12.0 & 300 & 1999 Apr 08 &	1.493	&	0.070	&	64.9	&	352	&	0.703	&	0.004	&	128.5	&	1.998	&	0.070	&	56.8	&	1	\\
HD 94509 & 9.1	& 120 & 1999 Apr 08 &	0.688	&	0.015	&	123.1	&	322	&	0.480	&	0.007	&	143.1	&	0.445	&	0.017	&	101.1	&	2	\\
HD 95881 & 8.3	& 10 & 1999 Jan 21 &	1.504	&	0.034	&	116.2	&	81	&	1.500	&	0.021	&	122.6	&	0.335	&	0.040	&	74.7	&	2	\\
PDS 327	& 8.5 & 12 & 1999 Jan 21 &	0.613	&	0.032	&	114.0	&	276	&	0.428	&	0.019	&	130.4	&	0.343	&	0.037	&	92.7	&	1	\\
HD 97048 & 8.5	& 10 & 1999 Jan 22 &	2.519	&	0.044	&	143.3	&	3	&	3.629	&	0.066	&	138.1	&	1.238	&	0.079	&	37.5	&	2	\\
PDS 339	& 7.8 & 7 & 1999 Jan 21 &	0.056	&	0.040	&	99.9	&	114	&	0.261	&	0.047	&	95.0	&	0.206	&	0.062	&	3.6	&	1	\\
PDS 340	& 6.8 & 3 & 1999 Jan 21 &	0.236	&	0.039	&	50.6	&	24	&	3.204	&	0.127	&	126.0	&	3.412	&	0.133	&	36.9	&	 1,2	\\
PDS 057	& 9.2 & 40 & 1999 Jan 21 &	0.737	&	0.027	&	90.9	&	488	&	1.122	&	0.017	&	89.3	&	0.388	&	0.032	&	176.2	&	 1,2	\\
PDS 344	& 13.2 & 300 & 1999 Apr 11 &	1.590	&	0.053	&	56.4	&	304	&	1.681	&	0.019	&	89.4	&	1.783	&	0.056	&	26.7	&	1	\\
PDS 061	& 6.6 & 2 & 1999 Jan 22 &	0.032	&	0.063	&	167.2	&	20	&	1.699	&	0.086	&	116.7	&	1.705	&	0.107	&	26.1	&	 1,2	\\
PDS 140	& 13.1 & 300 & 1999 Apr 11 &	1.852	&	0.099	&	89.4	&	476	&	0.773	&	0.008	&	83.6	&	1.106	&	0.099	&	93.4	&	1	\\
PDS 353	& 13.2 & 480 & 2000 Jun 21 &	0.666	&	0.056	&	22.5	&	1034	&	1.733	&	0.004	&	84.1	&	2.170	&	0.056	&	1.5	&	1	\\
Hen 3-847 & 10.6 & 80 & 1999 Apr 11 	&	0.348	&	0.035	&	20.7	&	186	&	0.631	&	0.017	&	53.0	&	0.575	&	0.039	&	159.6	&	2	\\
PDS 361	& 12.9 & 600 & 2000 Jun 22 & 0.219	&	0.034	&	1.3	&	1109	&	0.398	&	0.004	&	97.5	&	0.614	&	0.034	&	5.3	&	1	\\
PDS 364	& 13.5 & 300 & 1999 Apr 11 &	2.298	&	0.030	&	68.8	&	411	&	2.600	&	0.006	&	70.3	&	0.327	&	0.031	&	171.1	&	1	\\
PDS 067	& 13.5 & 600 & 2000 Jun 21 &	0.768	&	0.067	&	7.4	&	666	&	2.287	&	0.004	&	66.3	&	2.731	&	0.067	&	163.5	&	1	\\
PDS 069	& 9.8 & 60 & 1999 Feb 12 &	0.681	&	0.040	&	147.0	&	22	&	0.245	&	0.055	&	112.9	&	0.632	&	0.068	&	157.5	&	1	\\
HD 130437 & 10.0 & 120 & 1999 Apr 10	&	5.818	&	0.085	&	56.9	&	20	&	1.252	&	0.034	&	61.6	&	4.587	&	0.092	&	55.6	&	2	\\
HBC 596	& 12.8	& 300 & 1999 Apr 11 & 4.198	&	0.181	&	46.8	&	115	&	1.482	&	0.015	&	59.8	&	2.939	&	0.182	&	40.4	&	2	\\
HD 132947 & 8.9	& 15 & 1999 Apr 07 &	1.254	&	0.042	&	56.1	&	62	&	1.293	&	0.016	&	59.9	&	0.173	&	0.045	&	6.5	&	2	\\
PDS 389	& 14.2 & 300 & 1999 Jul 28 &	4.630	&	0.218	&	133.4	&	54	&	0.708	&	0.018	&	56.4	&	5.275	&	0.219	&	135.0	&	1	\\
PDS 394	& 13.5 & 300 & 1999 Jul 27 &	2.235	&	0.166	&	15.0	&	312	&	1.448	&	0.010	&	50.9	&	2.252	&	0.166	&	176.1	&	1	\\
PDS 395	& 8.4 & 10 & 1999 Apr 07 &	0.062	&	0.110	&	147.6	&	100	&	1.755	&	0.099	&	43.1	&	1.808	&	0.148	&	133.7	&	1	\\
PDS 144	& 12.8 & 300 & 1999 Apr 10 &	4.653	&	0.070	&	124.1	&	108	&	0.759	&	0.013	&	116.0	&	3.930	&	0.071	&	125.6	&	1	\\
PDS 398	& 7.1 & 3 & 1999 Nov 07 &	0.680	&	0.027	&	86.7	&	9	&	2.665	&	0.090	&	82.4	&	1.995	&	0.094	&	170.9	&	 1,2	\\
PDS 399	& 8.6 & 15 & 1999 Apr 10 &	1.705	&	0.022	&	61.0	&	385	&	1.553	&	0.018	&	54.9	&	0.378	&	0.028	&	91.2	&	1	\\
PDS 076	& 8.7 & 10 & 1999 Apr 07 &	0.803	&	0.039	&	81.5	&	20	&	0.077	&	0.027	&	85.6	&	0.727	&	0.047	&	81.1	&	 1,2	\\
PDS 406	& 13.9 & 300 & 1999 Apr 10 &	4.719	&	0.081	&	33.6	&	348	&	2.049	&	0.009	&	17.1	&	3.201	&	0.081	&	43.8	&	1	\\
PDS 078	& 8.2 & 10 & 1999 Apr 07 &	0.367	&	0.059	&	14.5	&	65	&	0.455	&	0.047	&	16.4	&	0.092	&	0.075	&	114.1	&	 1,2	\\
HD 144668 & 7.0	& 6 & 1999 Apr 08	& 0.579	&	0.019	&	166.5	&	7	&	0.498	&	0.010	&	5.6	&	0.361	&	0.021	&	137.2	&	2	\\
PDS 080	& 9.1 & 14 & 1999 Apr 07 &	0.039	&	0.052	&	45.8	&	47	&	0.589	&	0.059	&	4.9	&	0.585	&	0.079	&	93.0	&	1	\\
PDS 415	& 12.0 & 200 & 1999 Apr 08 &	1.418	&	0.022	&	28.5	&	59	&	1.536	&	0.019	&	29.7	&	0.133	&	0.029	&	132.9	&	1	\\
Hen 3-1191 & 13.7 & 300 & 1999 Apr 10	&	5.885	&	0.146	&	46.0	&	546	&	0.752	&	0.006	&	43.0	&	5.138	&	0.146	&	46.4	&	2	\\
HD 150193 & 8.9	& 15 & 1999 Apr 07 &	4.780	&	0.108	&	56.7	&	34	&	4.330	&	0.057	&	56.1	&	0.460	&	0.122	&	62.4	&	2	\\
PDS 431	& 13.4 & 300 & 1999 Jul 28 &	1.276	&	0.175	&	38.4	&	222	&	1.086	&	0.007	&	35.3	&	0.229	&	0.175	&	53.8	&	1	\\
V921 Sco & 11.4	& 80 & 1999 Apr 07 &	2.509	&	0.256	&	114.7	&	133	&	1.343	&	0.016	&	121.7	&	1.249	&	0.256	&	107.2	&	2	\\
KK Oph	& 11.9\tablenotemark{4} & 45 & 1999 Apr 10 &	3.426	&	0.092	&	169.6	&	62	&	0.913	&	0.063	&	33.3	&	3.505	&	0.112	&	162.1	&	2	\\
PDS 453	& 12.9 & 200 & 1999 Apr 07 &	3.608	&	0.060	&	48.9	&	412	&	2.635	&	0.021	&	7.3	&	4.208	&	0.064	&	68.1	&	1	\\
PDS 095	& 11.0 & 100 & 1999 Apr 11 &	1.634	&	0.096	&	34.3	&	70	&	1.041	&	0.022	&	165.6	&	2.048	&	0.098	&	49.4	&	1	\\
PDS 096	& 11.0 & 100 & 1999 Apr 11 &	1.662	&	0.031	&	175.3	&	95	&	1.329	&	0.011	&	177.5	&	0.352	&	0.033	&	166.9	&	1	\\
PDS 465\tablenotemark{5}	& 12.9 & 300 & 1999 Apr 11 &	9.152	&	0.251	&	43.4	&	501	&	1.298	&	0.009	&	48.3	&	7.876	&	0.251	&	42.6	&	1	\\
PDS 469	& 12.8 & 300 & 1999 Apr 11 &	1.315	&	0.031	&	72.4	&	637	&	1.739	&	0.007	&	62.7	&	0.663	&	0.032	&	132.1	&	1	\\
PDS 473	& 6.9 & 4 & 1999 Apr 08 &	0.417	&	0.035	&	36.0	&	63	&	1.140	&	0.063	&	42.5	&	0.740	&	0.072	&	136.1	&	 1,2	\\
PDS 477	& 14.4 & 300 & 1999 Jul 28 &	1.188	&	0.090	&	24.8	&	196	&	0.900	&	0.012	&	66.1	&	1.395	&	0.091	&	4.9	&	1	\\
PDS 514	& 8.2 & 12 & 1999 Apr 08 &	0.095	&	0.029	&	105.2	&	125	&	0.892	&	0.024	&	100.6	&	0.798	&	0.038	&	10.1	&	1	\\
PDS 518	& 12.2 & 300 & 1999 Jul 28 &	1.816	&	0.040	&	93.6	&	4	&	3.051	&	0.089	&	43.7	&	3.807	&	0.098	&	119.6	&	 1,2	\\
VV Ser	& 11.6\tablenotemark{4} & 400 & 2000 Jun 21 &	1.780	&	0.043	&	77.5	&	5	&	1.566	&	0.079	&	54.5	&	1.322	&	0.090	&	106.7	&	2	\\
MWC 300	& 10.5 & 300 & 2000 Jun 21 &	4.843	&	0.035	&	58.2	&	265	&	4.001	&	0.012	&	55.2	&	0.960	&	0.037	&	71.1	&	2	\\
PDS 520	& 14.7 & 300 & 1999 Jul 28 &	3.513	&	0.107	&	15.2	&	9	&	0.391	&	0.029	&	70.1	&	3.664	&	0.111	&	12.3	&	1	\\
HBC 284/1 & 12.5 & 360 & 2000 Jun 22 	&	0.916	&	0.095	&	67.1	&	501	&	0.637	&	0.009	&	56.8	&	0.390	&	0.095	&	84.6	&	2	\\
HBC 284/2 & 12.5 & 360 & 2000 Jun 22 	&	0.691	&	0.114	&	72.6	&	501	&	0.637	&	0.009	&	56.8	&	0.365	&	0.114	&	105.6	&	2	\\
PDS 530	& 14.0 & 300 & 1999 Jul 28 &	12.305	&	0.650	&	55.4	&	186	&	1.262	&	0.019	&	50.8	&	11.061	&	0.650	&	55.9	&	1	\\
PDS 543	& 12.5 & 360 & 2000 Jun 21 &	1.105	&	0.025	&	136.8	&	35	&	0.610	&	0.029	&	175.4	&	1.138	&	0.038	&	121.0	&	1	\\
PDS 545	& 8.8 & 18 & 1999 Apr 08 &	3.036	&	0.025	&	78.6	&	44	&	1.129	&	0.036	&	74.5	&	1.925	&	0.044	&	81.0	&	1	\\
PDS 551	& 16.6 & 600 & 2000 Jun 22 &	10.930	&	0.629	&	117.7	&	30	&	0.468	&	0.029	&	69.2	&	10.997	&	0.630	&	118.9	&	1	\\
PDS 564	& 7.4 & 4 & 1999 Apr 08 &	0.481	&	0.034	&	103.7	&	43	&	0.471	&	0.080	&	4.4	&	0.939	&	0.087	&	99.1	&	 1,2	\\
WW Vul	& 10.5 & 100 & 2000 Jun 21 &	0.849	&	0.070	&	151.3	&	336	&	2.065	&	0.011	&	12.2	&	2.118	&	0.071	&	113.9	&	2	\\
PDS 581\tablenotemark{5}	& 11.7 & 80 & 1999 Apr 07 &	12.427	&	0.145	&	44.4	&	772	&	0.326	&	0.008	&	69.5	&	12.221	&	0.145	&	43.8	&	1	\\
HD 190073 & 7.8	& 10 & 1998 Nov 25 &	0.377	&	0.026	&	94.5	&	15	&	0.439	&	0.054	&	97.9	&	0.079	&	0.060	&	25.2	&	2	
\enddata
\label{tab_pol}
\tablenotetext{1}{Magnitudes are from \citet{tor99} or SIMBAD.}
\tablenotetext{2}{The references of the last column are: (1) \citet{vie03}; (2) \citet{the94}; (3) \citet{tor99}.}
\tablenotetext{3}{The images of the field containing HD23302 and HD23480 do not include objects with signal high enough to estimate the foreground polarization.}
\tablenotetext{4}{B band magnitude.}
\tablenotetext{5}{PDS465 and PDS581 are post-AGB objects.}
\end{deluxetable}


\begin{thebibliography}{}

\bibitem[Bastien \& Menard(1990)]{bm90} Bastien, P., \& Menard, F.\ 1990, \apj, 364, 232 

\bibitem[Brown \& McLean(1977)]{bm77} Brown, J.~C., \& McLean, I.~S.\ 1977, \aap, 57, 141 

\bibitem[Cohen et al.(1984)]{coh84} Cohen, R.~J., Rowland, P.~R., \& Blair, M.~M.\ 1984, \mnras, 210, 425 

\bibitem[Davis \& Greenstein(1951)]{dav51} Davis, L. \& Greenstein, J.~L.\ 1951,
ApJ, 114, 206

\bibitem[Dyck \& Lonsdale(1979)]{dyc79} Dyck, H.~M., \& Lonsdale, C.~J.\ 1979, \aj, 84, 1339 

\bibitem[Gregorio-Hetem et al.(1992)]{gre92} Gregorio-Hetem, J., Lepine, J.~R.~D., Quast, G.~R., Torres, C.~A.~O.,  \& de La Reza, R.\ 1992, \aj, 103, 549 

\bibitem[Heckert \& Zeilik(1981)]{hec81} Heckert, P.~A., \& Zeilik, M., II 1981, \aj, 86, 1076 

\bibitem[Hodapp(1984)]{hod84} Hodapp, K.-W.\ 1984, \aap, 141, 255 

\bibitem[Jones \& Spitzer(1967)]{jon67} Jones, R.~V., \& Spitzer, L.~J.\ 1967, \apj, 147, 943 

\bibitem[Kobayashi et al.(1978)]{kob78} Kobayashi, Y., Kawara, K., Maihara, T., Okuda, H., Sato, S., \& Noguchi, K.\ 1978, \pasj, 30, 377 

\bibitem[Lazarian(2007)]{laz07} Lazarian, A.\ 2007, Journal of Quantitative Spectroscopy and Radiative Transfer, 106, 225 

\bibitem[Lazarian \& Hoang(2007)]{lh07} Lazarian, A., \& Hoang, T.\ 2007, \mnras, 378, 910

\bibitem[Magalh\~aes et al.(1996)]{mag96}Magalh\~aes, A.~M., Rodrigues, C.~V.,
Margoniner, V.~E., Pereyra, A.,  \& Heathcote, S., 1996, in ASP Conf. Ser. 97,
Polarimetry of the Interstellar Medium, Eds. W.~G. Roberge \& D.~C.~B. Whittet 
(San Francisco:ASP), 118

\bibitem[McKee \& Ostriker(2007)]{mo07} McKee, C.~F., \& Ostriker, E.~C.\ 2007, \araa, 45, 565 

\bibitem[M{\'e}nard \& Duch{\^e}ne(2004)]{men04} M{\'e}nard, F., \& Duch{\^e}ne, G.\ 2004, \aap, 425, 973 

\bibitem[Mouschovias \& Ciolek(1999)]{mou99} Mouschovias, T.~C., \& Ciolek, G.~E.\ 1999, NATO ASIC Proc.~540: The Origin of Stars and Planetary Systems, 305

\bibitem[Paltani(2004)]{pal04} Paltani, S.\ 2004, \aap, 420, 789 

\bibitem[Pereyra(2000)]{per00}Pereyra, A., 2000. Dust and Magnetic Fields in
Dense Regions of the Interstellar Medium, PhD Thesis, Univ. S\~ao Paulo

\bibitem[Pereyra \& Magalh{\~a}es(2002)]{per02} Pereyra, A., \& Magalh{\~a}es, A.~M.\ 2002, \apjs, 141, 469 

\bibitem[Strom \& Strom(1987)]{str87} Strom, S.~E., \& Strom, K.~M.\ 1987, Star Forming Regions, 115, 255 

\bibitem[Tamura \& Sato(1989)]{tam89} Tamura, M., \& Sato, S.\ 1989, \aj, 98, 1368 

\bibitem[Th\'e et al.(1994)]{the94} Th\'e, P.~S., de Winter, D., \& Perez, M.~R.\ 1994, \aaps, 104, 315 

\bibitem[Torres(1999)]{tor99} Torres, C.~A.~O.\ 1999, Special Publ. 10 (Rio de Janeiro Observatório Nacional)

\bibitem[Torres et al.(1995)]{tor95} Torres, C.~A.~O., Quast, G., de La Reza, R., Gregorio-Hetem, J., \& L\'epine, J.~R.~D.\ 1995, \aj, 109, 2146 

\bibitem[Vieira et al.(2003)]{vie03} Vieira, S.~L.~A., Corradi, W.~J.~B., Alencar, S.~H.~P., Mendes, L.~T.~S., Torres, C.~A.~O., Quast, G.~R., Guimar{\~a}es, M.~M., \& da Silva, L.\ 2003, \aj, 126, 2971 

\bibitem[Wade et al.(2009)]{wad09} Wade, G.~A., Alecian, E., Grunhut, J., Catala, C., Bagnulo, S., Folsom, C.~P., \& Landstreet, J.~D.\ 2009, arXiv:0901.0347 

\bibitem[Whitney \& Hartmann(1993)]{wh93} Whitney, B.~A., \& Hartmann, L.\ 1993, \apj, 402, 605 

\end{thebibliography}
\end{document}